\documentstyle[aps,prd,preprint,tighten,floats,epsf]{revtex}

\begin{document}
%

\preprint{PSU/TH/167}


\title{Thermal Properties of an Inflationary Universe}

\author{Arjun Berera}

\address{
   Department of Physics,
   Pennsylvania State University,
   University Park, PA 16802, U.S.A.
}

\maketitle

\begin{abstract}
An energetic justification of a thermal component during inflation is
given.  The thermal component
can act as a heat reservoir which induces thermal fluctuations
on the inflaton field system.  We showed previously that such
thermal fluctuations could dominate quantum fluctuations in producing
the initial seeds of density perturbations.  A Langevin-like
rate equation is derived from quantum field theory
which describes the production of
fluctuations in the inflaton field when acted upon by a simple
modeled heat reservoir.
In a certain limit this equation is shown to reduce to
the standard Langevin equation, which we used to construct "Warm Inflation"
scenarios in previous work.  A particle physics interpretation
of our system-reservoir model is offered.
\end{abstract}


\medskip

In press Phys. Rev. D 1996

\medskip

hep-th/9601134

\eject


\section{Introduction}
\label{sec.int}

According to inflationary cosmology, the large scale structure of the
present day universe is essentially a kinematic outcome of
exponential amplification of perturbing seeds in an initially smooth
universe \cite{guth,guthpi,brand}.
The differential microwave radiometer (DMR) on the Cosmic
Background Observer (COBE) has made the first direct probe of
the initial density perturbations through detection of
the temperature anisotropies in the cosmic background radiation (CBR).
These results are consistent with the scaling spectrum given by the inflation
model.  They also reinforce previously
known measurements, although done by less direct methods, that
the amplitude of
initial perturbations is
\begin{equation}
\Delta ({\bf k}) \equiv \frac{\delta \rho ({\bf k})}{\rho} = 10^{-3} - 10^{-5}
\label{amp}
\end{equation}
and approximately constant
for all wavenumber ${\bf k}$
\cite{smoot,benn}.  Here $\rho$ is the energy density in the present universe
and $\delta \rho ({\bf k})$ is its rms-deviation at wavenumber ${\bf k}$.

We can understand the underlying kinematic origin of large-scale
structure formation through a picture.
Let us imagine observing the universe before inflation.  It
is a small patch that is growing at light speed.  The process
of inflation can be thought of as a rapid stretching of this
patch in all directions.  Comoving and physical coordinates are
useful for further description.  Comoving coordinates
stretch with the patch so
do not change for points that are stationary with respect to
the patch.  Physical coordinates express distances in
terms of a physical measure, such as the local speed
of light.  A physical coordinate system is defined locally
to a given point on the patch.  It is useful sometimes
to understand global distances in terms of physical units.
For definitiveness,
at the onset of inflation, let the comoving coordinates
coincide with the physical coordinates.
If we imagine the
universe to be spherical with radius $R_0$ at the onset of inflation,
then during inflation, in comoving units,
the radius remains the same whereas in physical
units it grows as $e^{Ht}R_0$.
Thus during inflation
points on the preinflationary patch that are stationary in
the comoving frame
will rapidly move apart in terms of physical coordinates.

Identify a point {\bf a} on the preinflationary patch.
Suppose that before inflation instruments are placed at ${\bf a}$,
which maintain communication with all points
within reach by light since point ${\bf a}$
was created in the Big Bang.
As inflation begins, communication is first lost with
points furthest away from ${\bf a}$.  Immediately after inflation, only
points that were initially very close to ${\bf a}$ will remain in
communicative contact.  To be definitive, just before
inflation let the most distant points communicating with ${\bf a}$
be some distance $L_0 \sim 1/H$.  Here $H$ is the Hubble constant
with
$1/H = 5 \times 10^{-11} \ {\rm GeV}^{-1} = 10^{-23} \ {\rm cm}=
3 \times 10^{-35} \ {\rm sec}$. The time interval $1/H$ is
typically referred to as an e-fold.

To understand the behavior of physical measures during inflation,
let the origin of cosmic time, t=0, correspond to the
beginning of inflation.  The physical distance of a comoving
interval $\Delta x$ at time t will then be $e^{Ht} \Delta x_{phys}$.
On the other hand, for a light signal emitted at $t=0$, computing
along its geodesic $ds^2=0$, it will travel a physical distance
$(e^n/H) (1-e^{-n})$ after n e-folds of inflation.

The implication of these two relations to physical correlations
can be understood from the following example.
For a signal emitted at $t=0$,
points fixed with respect to the comoving coordinates (comoving
points) that are less than a comoving distance $(1/H)(1-e^{-1})$
will receive this signal before the first e-fold of inflation.
However comoving points greater than a comoving distance $1/H$
from ${\bf a}$ can never receive the light signal within
the inflation stage.  Only a sufficient time after inflation
could such points communicate with ${\bf a}$.  In terms
of physical distances, one concludes that comoving points
that are greater than a physical distance $1/H$ at $t=0$
will lose communicative contact with ${\bf a}$ all during
inflation.
Since nothing is special
about point ${\bf a}$ to any other point on the patch,
on the large scale physical distances at $t=0$ that are
greater than $1/H$ will act incoherently
all during inflation.
Thus correlations larger than $1/H$ physical units are thereafter
"frozen".   This phenomenon is sometimes referred to as
freeze-out or horizon crossing \cite{brand}.

If ${\bf a}$ emits a second light signal at the
end of the first e-fold, by the end of the
second e-fold this signal again will have traveled a physical
distance $(e/H)(1-e^{-1})$.  However comoving coordinates will
have stretched by a factor $e^2$
in terms of physical units. Thus only those points which are a comoving
distance $e^{-1}/H$ or less can receive the second signal at
sometime within the inflation stage.  Expressed in physical units,
the conclusion stated above from the first e-fold repeats itself for the
second e-fold:
comoving points
that are greater than a physical distance $1/H$ at $t=1/H$ hereafter
will lose communicative contact with ${\bf a}$ all during the rest of
inflation.
Finally a signal emitted from ${\bf a}$ at the end of
the n-th e-fold can be received
by comoving points within a comoving distance $e^{-n}/H$ from
${\bf a}$.  Thus comoving points greater than a physical distance $1/H$
at $t=n/H$ thereafter lose communicative contact with ${\bf a}$
during the rest of inflation.

Any local energy perturbations during inflation can affect a region of
characteristic physical length 1/H or less.
The earlier a given perturbation occurs,
the more elongated it will become due to inflation.  As such the largest
scales of energy density fluctuations in the post-inflationary
universe arose from the earliest perturbations during inflation.

In the standard scenario \cite{kolb}, inflationary expansion is assumed to
occur within a supercooled universe, in which
the initial energy density perturbations were produced
by quantum fluctuations.  However energetics does not
require a supercooled state.  Furthermore, attempts based on
this assumption have shown unnatural features.

For inflation, naturality has played an important role.  This
is understandable since
for phenomena that can not be directly observed, one attempts
a description starting with the most natural expectations.
The importance of naturality principles
is to provide guidance from more familiar
analogies with the hope of gaining predictability.
For inflation we can understand naturality as both a macroscopic and
microscopic one.   Macroscopically we would like a
description that rests with common-day experience.
Microscopically it should be consistent with the standard
model of particle physics.

Under both categories, the standard scenario has shown unsatisfactory
features \cite{guthpi,brand,mwu}.  Microscopically  slow-roll scalar field
dynamics requires an ultra-flat potential, although
no such potential is required otherwise for particle phenomenology.
Macroscopically, reheating requires globally coherent radiation
waves on the scale of the inflated universe.
Local incoherent heat transfer is more familiar to experience.
Furthermore, a globally coherent heating process requires a large scale
radiator, which in the standard scenario is the inflaton.
This raises the question about how the
random inflaton field configuration before and during inflation
attains quantum coherence at the end?

We can accept that naturality principles for inflation do not have
rigorous justifications and therefore
can be abandoned or softened.
However with this, any picture of inflation based on
familiar analogies
would require modifications. This loosens theoretical constraints
which otherwise are required to be
consistent with only the limited data from observation.
As such predictability
from theory becomes less definitive.  Nevertheless,
if that is the way nature works, then that is the way it is.

This would be an acceptable conclusion once all attempts for
a natural explanation have been examined.
If we digress back to this elementary point, we can rethink about
the known ways to induce energy density perturbations.  More general
to quantum fluctuations are thermal fluctuations.
We showed in \cite{bf1} that in the
context of near equilibrium thermodynamics,
during inflation thermal fluctuations could dominate quantum fluctuations
under certain conditions.  Neither energetics nor standard model
dynamics precludes a thermal state during inflation.
In \cite{wi} a model "Warm Inflation" scenario was
considered for a self interacting scalar field.  The solution
had the desirable feature of satisfying observational
constraints with minimal specifications placed
on the field theory.  The condition for slow-roll
was shifted from requirements on the potential to
a frictional force term that coupled the inflaton
to a thermalized heat reservoir.  By shifting to a frictionally
produced slow-roll, it also
gave a local heating mechanism.

The first goal of this paper is to clarify the energetics in a
thermal inflationary environment.  Secondly we derive from quantum
field theory a generalized Langevin equation for the scalar
inflaton field coupled to a modeled heat reservoir system.
In certain limits, which we state, the equation
reduces to the one we used in \cite{wi}.
Finally with these ingredients, an interpretation in terms of
particle physics is attempted.  To maintain physical
clarity, our derivation is Hamiltonian based
and performed in a cubic box with periodic boundary conditions.
Appendix A provides a convenient reference to thermal properties
of free fields that are used often in the text.  It also
relates our notation to standard form as well as shows how to
take the infinite volume limit.  Appendix B gives an alternative
derivation of the Langevin equation from that in section (III).  The
purpose for this is discussed in section (III).

\section{Energetics}
\label{sec.ener}

Let us account for the total energy in the inflationary universe.
Consider a scalar inflaton field with Hamiltonian density defined with respect
to the physical volume
\begin{equation}
{\cal H} = \frac{1}{2} {\dot \phi}^2 + V(\phi).
\end{equation}
First suppose that the inflaton is the only system in the universe.
If during inflation the inflaton has negligible kinetic energy,
$\langle {\dot \phi}^2 \rangle \approx 0$, then the
energy density during inflation would be all potential from $V(\phi_0)$.
Such a situation could occur if $\phi_0$ were at
a local extrema where $V'(\phi_0)=0$.  Although the energy
density remains constant in this case, the volume $U$ of the universe
in physical units would grow after
n e-folds to $U(n) \sim R_0^3 e^{3n}$ where $U(0) \sim R_0^3$ is
the initial volume of the universe just before inflation.
In terms of total energy, $E_T(n) \equiv U(n) <{\cal H}> $,
it would be after $n$ e-folds of inflation
\begin{equation}
E_T(n)=V(\phi_0) U(0)e^{3n}=E_T(0)e^{3n}.
\end{equation}

Turning to the more realistic situation in which there is
kinetic energy, it is conceptually helpful
to first understand energy transfer in the classical limit for
the inflaton field.  The quantum mechanical problem is
treated in section (III).  For the classical limit in the expanding
universe, the rate of change of the energy density
can be expressed as
\begin{equation}
\frac{d{\cal H}(t)}{dt}= -3H{\dot \phi}^2(t).
\label{disexp}
\end{equation}
If in addition the inflaton field expels energy to some other system, one
can express this, when treating the inflaton as an isolated system,
by adding a dissipative term to the right-hand-side
of eq. (\ref{disexp}).  If we choose the specific form
to have the lowest time derivative and be even in the field,
eq. (\ref{disexp}) becomes
\begin{equation}
\frac{d{\cal H}(t)}{dt}= -3H{\dot \phi}^2(t) -
\int dt' {\dot \phi}(t) \Gamma(t,t') {\dot \phi}(t').
\label{diseq}
\end{equation}
This implies that the equation of motion for the inflaton is
\begin{equation}
\ddot{\phi}(t) + 3H {\dot \phi}(t) + \int^t \Gamma (t,t') {\dot \phi}(t') dt'
+V'(\phi(t)) = 0.
\label{eqsmotion}
\end{equation}
In the next section we derive a quantum operator equation
similar to this but also include a random force term and treat
spatial variations.

Having inserted a dissipative term in eqs. (\ref{diseq}), energy balance
implies that there must be some other system receiving this energy.
If the second system is sufficiently large, it will
act as a heat reservoir which induces fluctuations
on the inflaton field.  In the next section we examine
a model heat reservoir system which
we assume
is thermalized.  For the reservoir,
we do not commit ourselves to a specific
particle physics realization.  However,
in section (IV)
a particle physics interpretation is offered.  Furthermore, one
may question the assumption of thermalization for the heat reservoir.
More general would be some other statistical
distribution.  However for the present work we assume that
the heat reservoir is in thermal equilibrium at some temperature T.
Further treatment of this problem would require details about
the dynamics beyond what we consider.
Finally during inflation, the temperature also could be a function of
cosmic time, but we consider it fixed.

What is not an assumption and the important point to establish here
is the energetic justification of the system - heat reservoir
decomposition.  The validity of this as well as consistency with
the inflation solution requires
\begin{equation}
\delta \rho_{\phi}({\bf x}, t) \ll \rho_r(t) \ll \rho_{\phi}(t).
\label{syshr}
\end{equation}
Here
\begin{equation}
\delta \rho_{\phi}({\bf x}, t)= {\cal H}(\phi({\bf x},t)) -{\cal H}(\phi_0(t))
=V'(\phi_0(t)) \delta \phi({\bf x}, t)
\end{equation}
is the energy density contained in the fluctuations of the inflaton
field
\begin{equation}
\delta \phi({\bf x},t) = \phi({\bf x},t)-\phi_0(t),
\end{equation}
$\rho_r(t)$ is the energy density in the heat reservoir,
and $\rho_{\phi}(t)$ is the vacuum energy density, with
all of these evaluated at cosmic time t.

The first requirement, from the right inequality in eq. (\ref{syshr}),
is the vacuum dominance
condition needed for inflation and the second requirement,
from the left inequality, is needed for
the system-heat reservoir decomposition.
In \cite{wi} the warm inflation scenarios that we found were for
\begin{equation}
R \equiv \frac{\rho_r}{\rho_{\phi}} \leq 10^{-2}.
\label{vacineq}
\end{equation}
This is sufficient to satisfy the first requirement of vacuum energy
dominance.  This statement can be strengthened for certain
solution regimes found, which were consistent with
observation and had $R$ one to two orders of magnitude less
than eq. (\ref{vacineq}).
The second requirement
can be established
from the observed amplitude in eq. (\ref{amp}) since \cite{brand,bard}
\begin{eqnarray}
10^{-3} - 10^{-5} \sim
\left(\frac{\delta \rho_{\phi}({\bf k})}{\rho}\right)_{HE}
=\left(\frac{\delta \rho_{\phi}({\bf k})}{{\dot \phi}^2+
\frac{4}{3} \rho_r}\right)_{HC}
\ll \left(\frac{\delta \rho_{\phi}({\bf k})}{\rho_{\phi}}\right)_{HC}.
\label{ampcomp}
\end{eqnarray}
Here $HC$ and $HE$ mean horizon crossing and entry respectively, and
${\bf k}$ is the comoving wavenumber with horizon crossing occurring
at cosmic time t such that $|{\bf k}| e^{-Ht} \sim H$.
On the left-hand-side $\delta \rho$ from eq. (\ref{amp})
has been equated to $\delta \rho_{\phi}$.
In \cite{bf1} we found  that the minimal thermodynamic
requirement for a thermal scenario was $\rho_r \sim {\dot \phi}^2$,
although for the model in \cite{wi}
for all cases $\rho_r \gg {\dot \phi}^2$.
In either case from eq. (\ref{ampcomp}) we have
\begin{equation}
\left(\frac{\delta \rho_{\phi}({\bf k})}{{\dot \phi}^2+ \frac{4}{3}
\rho_r}\right)_{HC}
\sim
\left(\frac{\delta \rho_{\phi}({\bf k})}{ \rho_r}\right)_{HC} \sim
10^{-3} - 10^{-5},
\end{equation}
which satisfies the second requirement in eq. (\ref{syshr}).
Thus energetically a thermal component can exist during inflation.

\section{Field Theory}
\label{sec.field}

In this section a field theory derivation of the operator
equation of motion for the inflaton is given
which has the form of a Langevin-like
rate equation.  In \cite{bf1,wi}
such a rate equation was postulated.  Below we consider a
simple model heat reservoir system which is coupled
linearly to the completely interacting inflaton field.  The total
system-reservoir
Lagrangian is
\begin{equation}
L_T=L_S+L_R+L_I,
\end{equation}
where on the right-hand-side the Lagrangians
are $L_S$ for the inflaton (system), $L_R$ for the reservoir,
and $L_I$ for the interaction between system and reservoir.
The inflaton's Lagrangian, $L_S$, can have an arbitrary
potential and accounts for the expansion term.
It has the familiar form
\begin{equation}
L_S= \int_V d^3{\bf x} e^{3Ht}\left[\frac{1}{2} \left(
(\partial_0 \phi({\bf x},t))^2-
(e^{-Ht}{\bf \nabla} \phi({\bf x},t))^2 - m^2 \phi^2({\bf x},t) \right)
-V(\phi({\bf x},t)) \right],
\label{langs}
\end{equation}
where the potential can have the general expansion
\begin{equation}
V(\phi) = \sum_{n=3}^{\infty} \frac{g_n}{n!} \phi^n({\bf x},t).
\end{equation}
In this paper we derive the effective operator equation of motion for
the inflaton, but we do not study it any further.  Thus
we will not address
the issues of renormalization, which would be connected with solving
this equation. As such, our formal derivation is valid whether
$L_S$ represents an elementary or effective field theory.

The heat reservoir Lagrangian, $L_R$, is modeled
as a set of free fields each characterized by a mass $\mu_i$.
It is written as
\begin{equation}
L_R= \sum_i \int_V d^3{\bf x} e^{3Ht} \frac{1}{2}
\left[ (\partial_0 \eta_i({\bf x},t))^2-
(e^{-Ht}{\bf \nabla} \eta_i({\bf x},t))^2 - \mu_i^2 \eta_i^2({\bf x},t)
\right].
\label{lres}
\end{equation}
Each reservoir field is coupled linearly to the
inflaton with a coupling constant $\alpha_i$ through
the interaction Lagrangian
\begin{equation}
L_I=-\sum_i \alpha_i \int_V d^3{\bf x}e^{3Ht} \eta_i({\bf x},t) \phi({\bf
x},t).
\label{resli}
\end{equation}
Note that the coupling constants $\{\alpha_i\}$ carry
engineering dimension 2.

The derivation given below treats the energy transfer between
the system and reservoir.
However the modeled reservoir in eq. (\ref{lres})
does not have internal interactions.
Furthermore energy transfer between the inflaton
and the reservoir is not fully treated.
To understand the latter two points, note that our problem has a difference
to standard problems in Langevin dynamics.  The reservoir
acts as a large system on the fluctuations of the inflaton field,
which is a standard situation for applying
Langevin dynamics.  However the difference is that
the vacuum energy of the inflaton, and particularly through
the zero mode, acts as a energy source for maintaining
the energy density of the reservoir, which otherwise
would diminish due to inflationary expansion.
This introduces two complications. The first concerns
thermalization for the reservoir.  In a standard Langevin
problem, it is assumed that the system will weakly interact
with the subsystems of the reservoir.  There can be many subsystems
so that the total effect of the reservoir on the system can be
strong.  However each subsystem of the reservoir is
affected only weakly by the system.  In such a circumstance the
issue of maintaining an initially thermalized state for the
reservoir is not acute. As such, internal reservoir interactions
are not crucial to know.
In our problem we can accept that the couplings
$\alpha_i$ in eq. (\ref{resli}) are small (although the derivation below
holds for arbitrary $\alpha_i$'s).
The question of concern is for whatever vacuum energy that
is transferred into the reservoir, can it thermalize on a time scale
shorter than 1/$H$?  Even if our modeled reservoir
had internal interactions, to prove thermalization from first principle
dynamics would be complicated  (some attempts in one-dimensional
models are given in \cite{odther}).

In our treatment we assume
that thermalization occurs and make appropriate
by-hand adjustments.  These are detailed in the
derivation, when relevant. In the next section we
will return to the elementary question of thermalization again.
Alongside with this problem is the second complication,
which is in regards to
energy transfer from the inflaton vacuum to the reservoir.
Our derivation accounts for only sub-horizon scale physics.
As such it treats the fluctuations induced on the modes
of the inflaton field by the reservoir,
while the modes are sub-horizon scale,
${\bf k}_{\rm physical} > H$.  However our derivation is not justified for
treating the zero mode interaction between the inflaton and the
reservoir, since that also involves super-horizon scale
physics.  The suggestive guess is that the "naive"
operator equation derived below for the zero mode,
but without a thermalization assumption on the super-horizon scale
modes of the reservoir, is approximately
valid. We will return to this issue in the conclusion.

It is worth noting that in certain limits
the thermal inflation problem reduces
to the standard Langevin problem, although this limit
did not prove useful in the warm inflation scenarios
we considered in \cite{wi}.  This limit
is given in Appendix B as well as the derivation of
the Langevin equation in this limit.

Our first goal is to derive the effective equation of motion for
the inflaton with the reservoir field variables eliminated.
The equation is valid for a time interval $|\Delta t|< H^{-1}$.
We will derive the equation for an arbitrary comoving mode
of the inflaton field and for the m-th time interval,
$t_m$ to $t_{m+1}$,  with
m arbitrary, where $t_m \equiv m \Delta t$ and $t_0=0$
is fixed as the starting time of inflation.
A complete solution for the inflaton's evolution
can be obtained by piecewise construction over all time intervals.
The formal derivation is not different for
physical wave-numbers that are sub-horizon or super-horizon
scale, although as stated above,  the approximations leading to
the derivation are valid only for the former.

In solving
the equations of motion for the reservoir fields, three approximations
are made in the m-th time interval for
every m.  Firstly
the red-shifting factor of a given comoving wavenumber
${\bf k}_n$ is held fixed at $e^{-2Ht_m}$.
Secondly, at the beginning of every time interval, the state of the
reservoir field operators is readjusted. Finally
the uncoupled modes of the reservoir fields are assumed to
obey a canonical distribution with respect to their physical
frequency
\begin{equation}
\omega^i_{n(t_m)} \equiv e^{-2Ht_m}{\bf k}^2_n +\mu_i^2.
\label{omegat}
\end{equation}
Here time t has been demoted into the subscript to
signify that it is treated as an adiabatic parameter as
far as the wavenumber is concerned.

The first approximation is made to simplify the calculation
so that it can be solved analytically.  It can be dropped if
one is willing to apply more sophisticated methods of solution.
The latter two approximations are physically
motivated.  They are by-hand treatments of the interactions
amongst the reservoir fields.  Details about the second
approximation are given within the derivation when relevant.
The third approximation implements our thermalization assumption
for the reservoir.  In the derivation we are careful to separate these
two approximations.  The second is made at the operator level.
The third is a statement about the state.  For it, we are firstly
assuming that the description of the reservoir is statistical and
secondly that the particular distribution is canonical.
In the next section we argue that the reservoir state is created
from quantum decay processes, in which case a statistical
description is inherently required.  The assumption of
being a canonical distribution seems the most obvious first guess.

The final equation of motion for the inflaton
will be stochastic since the reservoir
state is specified by a statistical distribution.
The final equation of motion for the inflaton
superficially will appear nonconservative since the reservoir fields
are going to be eliminated.  Under certain conditions placed on the reservoir
Hamiltonian, we derive the Langevin equation used in \cite{wi}.
This limit is examined and we verify the fluctuation-dissipation
theorem in its standard form \cite{cw}.

The derivation below follows well known methods from
non-equilibrium statistical mechanics which have been refined
over the years \cite{kac,sen,nag,mcl}.  A primary motivation
for this ongoing effort has been to understand the universal
properties of the Langevin equation and to obtain a possible explanation
from first principles \cite{kac,kacb}.
This believed universality is one reason for us to start our
study of stochastic dynamics for thermal scenarios
with the Langevin equation.
Our derivation below follows closest to \cite{kac} and the model
for the heat reservoir follows \cite{mcl}.
We have made some modifications to these works, which were
for quantum mechanical models, in order that we can treat
a quantum field and account for the
expansion term.
In Appendix B there is an alternative derivation which
obtains the rate equation in certain limits for an arbitrary reservoir
Hamiltonian.

Before proceeding, Let us review literature that is related to the
present work.
Applications in cosmology using Langevin dynamics
have been done in \cite{otherc,otherc2}, although
both our methods and motivation differ from
these works. Studies of finite temperature field theory in Robertson-Walker
Universes have been done in \cite{hu1}.  Finally
a calculation with similar objectives to ours in this paper
is given in \cite{cornbru}.  There a path integral derivation is
presented of the inflaton's evolution when coupled to a thermal
bath in de Sitter space.  However the authors did not
completely examine the dissipative properties of such a system.
As such they apparently missed the connection to warm inflation type
scenarios, which  for us is the starting motivation to the present
formal exercise.

We perform our derivation in a cube for the three spatial directions,
which is centered at the origin with sides at $\pm L/2$
in each direction.  The Fourier expansion of a generic field
is from eqs. (\ref{fourchi})
\begin{equation}
\chi({\bf x},t)=\frac{1}{L^3} \sum_n \chi({\bf k}_n,t)
e^{i{\bf k}_n \cdot {\bf x}},
\label{xchi}
\end{equation}
where throughout this paper we use the notation
${\bf k}_n \equiv 2 \pi {\bf n}/L$, with ${\bf n} \equiv (n_x,n_y,n_z)$
and
\begin{equation}
\sum_n \equiv \sum_{n_x=-\infty}^{\infty}
\sum_{n_y=-\infty}^{\infty}
\sum_{n_z=-\infty}^{\infty}.
\end{equation}
The argument of the coordinate (${\bf x}$) and
the momentum (${\bf k}$) space fields are always given to distinguish the
two.

The system-reservoir Lagrangian in terms of the Fourier modes
as defined in eq. (\ref{xchi}) is $L_T=L_S+L_R+L_I$, where
\begin{equation}
L_S=\frac{e^{3Ht}}{ L^3} \left\{ \frac{1}{2} \sum_n
\left[{\dot \phi}({\bf k}_n,t) {\dot \phi}(-{\bf k}_n,t)
-(e^{-2Ht}{\bf k}_n^2+m^2) \phi({\bf k}_n,t) \phi(-{\bf k}_n,t)\right]
-V_F \right\},
\end{equation}
\begin{equation}
L_R=\frac{e^{3Ht}}{L^3} \sum_i \sum_n \frac{1}{2}
\left[{\dot \eta}_i({\bf k}_n,t) {\dot \eta}_i(-{\bf k}_n,t)
-(e^{-2Ht}{\bf k}_n^2+\mu_i^2) \eta_i({\bf k}_n,t) \eta_i(-{\bf k}_n,t)
\right],
\end{equation}
and
\begin{equation}
L_I=-\frac{e^{3Ht}}{L^3}\sum_i \sum_n \alpha_i
\eta_i({\bf k}_n,t) \phi(-{\bf k}_n,t)
\end{equation}
with
\begin{equation}
V_F \equiv L^3 \int d^3{\bf x} V(\phi) = \sum_{n=3}^{\infty}
\frac{1}{L^{3n-6}} \sum_{m_1, \cdots, m_{n-1}}
\phi({\bf k}_{m_1},t) \cdots \phi({\bf k}_{m_{n-1}},t)
\phi(-{\bf k}_{m_1} - \cdots -{\bf k}_{m_{n-1}},t).
\end{equation}

The conjugate momentum to any scalar field $\chi({\bf k}_n,t)$
is
\begin{equation}
\pi_{\chi}({\bf k}_n,t) \equiv
\frac{\partial L}{\partial {\dot \chi}({\bf k}_n,t)}
=\frac{e^{3Ht}}{L^3} {\dot \chi}(-{\bf k}_n,t).
\end{equation}
Converting to the Hamiltonian
$H_T=\pi_{\phi} {\dot \phi} + \sum_i \pi_{\eta_i} {\dot \eta}_i -L_T$,
we obtain
\begin{equation}
H_S=\sum_n \frac{1}{2} \left[ e^{-3Ht} L^3
\pi_{\phi}({\bf k}_n,t)\pi_{\phi}(-{\bf k}_n,t)+ \frac{e^{3Ht}}{L^3}
(e^{-2Ht}{\bf k}_n^2+m^2)\phi({\bf k}_n,t)\phi(-{\bf k}_n,t)\right] +
\frac{e^{3Ht}}{L^3}V_F,
\label{hsys}
\end{equation}
\begin{equation}
H_R=\sum_i \sum_n \frac{1}{2} \left[e^{-3Ht} L^3
\pi_{\eta_i}(k_n,t)\pi_{\eta_i}(-{\bf k}_n,t)+ \frac{e^{3Ht}}{L^3}
(e^{-2Ht}{\bf k}_n^2+\mu_i^2)\eta_i({\bf k}_n,t)\eta_i(-{\bf k}_n,t) \right],
\label{hres}
\end{equation}
and $H_I=-L_I$.  Our notation is that all Hamiltonians
have some specifying subscript, which then leaves the Hubble
constant to be $H$.

To quantize the theory the postulated equal time commutation relations
(CCR) are given in eqs. (\ref{ccr}) and (\ref{fccr}).
The operator equations of motion from $H_T$ are
\begin{equation}
{\dot \phi}({\bf k}_n, t) = i[H_T,\phi({\bf k}_n,t)]=
e^{-3Ht}L^3 \pi_{\phi}({\bf k}_n,t),
\end{equation}
\begin{equation}
{\dot \pi}_{\phi} ({\bf k}_n, t) = i[H_T,\pi_{\phi}({\bf k}_n,t)]=
-\frac{e^{3Ht}}{L^3} \left[ (e^{-2Ht}{\bf k}_n^2+m^2) \phi({\bf k}_n,t)
+\frac{\delta V_F}{\delta \phi(-{\bf k}_n,t)}
+\sum_i \alpha_i \eta_i({\bf k}_n,t) \right],
\end{equation}
\begin{equation}
{\dot \eta}_i({\bf k}_n, t) =
e^{-3Ht}L^3 \pi_{\eta_i}({\bf k}_n,t),
\end{equation}
and
\begin{equation}
{\dot \pi}_{\eta_i} ({\bf k}_n, t) =
-\frac{e^{3Ht}}{L^3} \left[ (e^{-2Ht}{\bf k}_n^2+\mu_i^2) \eta_i({\bf k}_n,t)
+\alpha_i \phi({\bf k}_n,t) \right].
\end{equation}

The resulting second order field equations are
\begin{equation}
\ddot{\phi} ({\bf k}_n, t) +3H{\dot \phi}({\bf k}_n,t)
+[e^{-2Ht}{\bf k}_n^2+m^2]\phi({\bf k}_n,t)+
\frac{\delta V_F}{\delta \phi(-{\bf k}_n,t)}
+\sum_i \alpha_i \eta_i({\bf k}_n,t)=0
\label{phieq}
\end{equation}
and
\begin{equation}
\ddot{\eta}_i ({\bf k}_n, t) +3H{\dot \eta}_i({\bf k}_n,t)
+[e^{-2Ht}{\bf k}_n^2+\mu_i^2]\eta_i({\bf k}_n,t)
+ \alpha_i \phi({\bf k}_n,t)=0.
\label{eqnchi}
\end{equation}

We now implement our second approximation discussed
earlier.  At the beginning of every time interval
$t_m$, we readjust the state of the reservoir fields
as
\begin{equation}
\eta_i({\bf k}_n,t_m; t_{m-1}) \rightarrow
\eta^0_i({\bf k}_n,0; t_m),
\end{equation}
where $\eta_i({\bf k}_n,t;t_{m-1})$ and $\eta_i^0({\bf k}_n, t;t_m)$
are the solutions of eq. (\ref{eqnchi}) at time $t$ respectively
with and without the coupling term to $\phi$ and with the frequency
$\omega_{n(t)}^i$ held fixed at respectively $t_{m-1}$ and $t_m$.
Formally this operation can be viewed as a set
of impulsive forces that act on the reservoir.
The first purpose of these adjustments
is to add sufficient energy so that the reservoir's
energy density with respect to physical volume remains constant.
Secondly these adjustments are an external treatment of interactions
within the reservoir.
They shift each field back to its free field state
at $t=0$ but with its physical frequency decreased by a little.
This operation is in preparation for the thermalization
assumption we will make below on the reservoir's state.
The first purpose given above acts as a constraint on
the inflaton's evolution.  Consistency with the inflaton's
equation of motion for a particular model was demonstrated
in \cite{wi}.
The discretized treatment simplifies the
calculation.  One expects the actual dynamics to be smooth and continuous.

The solution for the oscillator field $\eta_i$ for the time interval
$[t_m, t_{m+1}]$ is
\begin{eqnarray}
\eta_i({\bf k}_n,t;t_m) &=& \eta^0_i({\bf k}_n,t-t_m;t_m)
-\frac{\alpha_i}{\Omega^{i2}_{n(t_m)}} \left[ \phi({\bf k}_n,t)
-\phi({\bf k}_n,t_m) e^{\frac{-3H(t-t_m)}{2}} \cos\Omega^i_{n(t_m)}(t-t_m)
\right.
\nonumber\\
&-& \left. e^{-\frac{3}{2}H(t-t_m)} \int_{t_m}^t dt'
\cos \Omega^i_{n(t_m)}(t-t') e^{\frac{3}{2}H(t'-t_m)}
\left({\dot \phi}({\bf k}_n,t')+\frac{3H}{2}\phi({\bf k}_n, t')\right)\right]
\label{etasol}
\end{eqnarray}
where
\begin{equation}
\Omega^i_{n(t)}=\sqrt{e^{-2Ht}{\bf k}_n^2+\mu_i^2-\frac{9H^2}{4}}
\label{Omega}
\end{equation}
and
\begin{equation}
\eta_i^0({\bf k}_n,t; t_m) =
e^{-\frac{3Ht}{2}}\left[\eta_i^0({\bf k}_n,0) \cos \Omega^i_{n(t_m)}(t)+
\frac{1}{\Omega^i_{n(t_m)}}\left(\frac{3H}{2}\eta_i^0({\bf k}_n,0)+
L^3 \pi_{\eta_i}({\bf k}_n,0)\right) \sin \Omega^i_{n(t_m)}(t)\right]
\label{etafree}
\end{equation}
is the solution for the free reservoir fields with the
time dependent physical frequency eq. (\ref{omegat})
of the comoving mode held
fixed at $t_m$.
As in eq. (\ref{omegat}), the
subscript $n(t)$ in eq. (\ref{Omega}) refers to the physical
wavenumber of the comoving mode ${\bf k}_n$
at cosmic time t.
Substituting eq. (\ref{etasol}) into eq. (\ref{phieq}),
we obtain for $t_m < t < t_{m+1}$
\begin{eqnarray}
\ddot{\phi}({\bf k}_n,t) &+&3H{\dot \phi}({\bf k}_n,t)
+ e^{-\frac{3}{2}H(t-t_m)}\sum_i \left( \frac{\alpha_i}{\Omega^i_{n(t_m)}}
\right)^2
\int_{t_m}^t dt' e^{\frac{3}{2}H(t'-t_m)} \cos \Omega^i_{n(t_m)}(t-t')
{\dot \phi}({\bf k}_n,t')
\nonumber\\
&+&
(e^{-2Ht}{\bf k}_n^2+m^2) \phi({\bf k}_n,t)
- \sum_i \left( \frac{\alpha_i}{\Omega^i_{n(t_m)}} \right)^2
\left[ \phi({\bf k}_n,t) - \phi({\bf k}_n,t_m)e^{-\frac{3}{2}H(t-t_m)}
\cos \Omega^i_{n(t)}(t-t_m) \right.
\nonumber\\
&-& \left.\frac{3H}{2}e^{-\frac{3}{2}H(t-t_m)}
\int_{t_m}^t dt' e^{\frac{3}{2}H(t'-t_m)} \cos \Omega^i_{n(t)}(t-t')
\phi({\bf k}_n,t')\right]
\nonumber\\
&+&\frac{\delta V_F (\phi)}{\delta \phi (-{\bf k}_n,t)}=\eta({\bf k}_n,t;t_m)
\label{rateeq}
\end{eqnarray}
where
\begin{equation}
\eta({\bf k}_n,t;t_m)=-\sum_i \alpha_i \eta_i^0({\bf k}_n,t-t_m;t_m).
\end{equation}

Up to this point, no statistical assumption has been made.
Following our earlier discussion, we now
assume that the free reservoir fields are canonically
distributed.  The statistical mechanics for the reservoir
system based on our above approximations is the same within
each e-fold as in flat-space.  For this Appendix A has been
provided as a useful reference.  As a clarification, no
approximation has been made in treating the inflaton system.
For the reservoir fields from eq. (\ref{chidist}) for $t_m< t,t' < t_{m+1}$,
we have
\begin{eqnarray}
\langle \langle \eta_i({\bf k}_n,t-t_m;t_m)&&\eta_i({\bf k}'_n,t'-t_m;t_m
\rangle \rangle_T =
\nonumber\\
&&\frac{L^3}{2} e^{-\frac{3H}{2}(t+t'-2t_m)} \delta_{n,-n'} \left[ \left(
\frac{1}{2 \omega_{n(t_m)}^i} \left[ \cos \omega^i_{n(t_m)}(t-t')
+\cos \omega^i_{n(t_m)}(t+t'-2t_m) \right] \right. \right.
\nonumber\\
&+& \left. \left.
\frac{1}{2\Omega^{i \ 2}_{n(t_m)}}\left( \frac{9H^2}{4 \omega^i_{n(t_m)}}
+ \omega^i_{n(t_m)} \right)
\left[ \cos \omega^i_{n(t_m)}(t-t')
-\cos \omega^i_{n(t_m)}(t+t'-2t_m) \right] \right. \right.
\nonumber\\
&+& \left. \left. \frac{3H}{2 \Omega^i_{n(t_m)} \omega^i_{n(t_m)}}
\sin \omega^i_{n(t_m)}(t+t'-2t_m) \right)
\coth\left( \frac{\omega^i_{n(t_m)}}{2T} \right) \right.
\nonumber\\
&-& \left.  \frac{i}{\Omega^i_{n(t_m)}}
\sin \omega^i_{n(t_m)}(t-t') \right].
\label{chiassump}
\end{eqnarray}
Eq. (\ref{rateeq}) with the assumption eq. (\ref{chiassump})
is the general rate equation and the main result of this
section.  For use later,
the symmetric correlation function
of the force operator is
\begin{eqnarray}
S_T({\bf k}_n,t,t';t_m) &\equiv& \frac{1}{2} \langle \langle
\eta({\bf k}_n,t;t_m)
\eta(-{\bf k}_n,t';t_m) +
\eta(-{\bf k}_n,t';t_m) \eta({\bf k}_n,t;t_m) \rangle \rangle_T
\nonumber \\
&=&\frac{L^3}{4} e^{-\frac{3H}{2}(t+t'-2t_m)} \sum_i \alpha_i^2 \left[
\frac{1}{\omega_{n(t_m)}^i} \left[ \cos \omega^i_{n(t_m)}(t-t')
+\cos \omega^i_{n(t_m)}(t+t'-2t_m) \right] \right.
\nonumber\\
&+& \left.
\frac{1}{\Omega^{i \ 2}_{n(t_m)}}\left( \frac{9H^2}{4 \omega^i_{n(t_m)}}
+ \omega^i_{n(t_m)} \right)
\left[ \cos \omega^i_{n(t_m)}(t-t')
-\cos \omega^i_{n(t_m)}(t+t'-2t_m) \right] \right.
\nonumber\\
&+& \left. \frac{3H}{\Omega^i_{n(t_m)} \omega^i_{n(t_m)}}
\sin \omega^i_{n(t_m)}(t+t'-2t_m) \right]
\coth\left( \frac{\omega^i_{n(t_m)}}{2T} \right).
\label{symmcor}
\end{eqnarray}

We next derive a limiting form of eq. (\ref{rateeq}). Consider
a large number of oscillator fields
$\eta_i$ that are represented by a continuous distribution
as
\begin{equation}
\sum_i \rightarrow \int d\mu N(\mu),
\label{spec}
\end{equation}
where $N(\mu)$ is the spectral weight function.
Assume that $N(\mu)$ is nonzero only in the
interval $\mu_l < \mu < \mu_u$ with
$\Delta_{\mu} \equiv \mu_u - \mu_l$ and the corresponding
definition $\Delta^{\Omega}_{n(t_m)} \equiv
\sqrt{e^{-2Ht}{\bf k}_n^2+\mu_u^2-\frac{9H^2}{4}}
-\sqrt{e^{-2Ht}{\bf k}_n^2+\mu_l^2-\frac{9H^2}{4}}$.
Also assume that
\begin{equation}
N(\mu) (\frac{\alpha(\mu)}{\Omega^{\mu}_{n(t_m)}})^2 = \frac{2 \Gamma}{\pi}
\label{specconst}
\end{equation}
is a constant over the interval $\Delta_{\mu}$.
If $\Delta^{\Omega}_{n(t_m)} \gg H$, then
for
$1/\Delta^{\Omega}_{n(t_m)} \ll t < 1/H$
\begin{equation}
\int_0^{\infty} d \mu N(\mu) (\frac{\alpha(\mu)}{\Omega^{\mu}_{n(t_m)}})^2
\cos \Omega^{\mu}_{n(t_m)} t = 2 \Gamma \delta_{1/\Delta^{\Omega}_{n(t_m)}}
(t).
\label{ndist}
\end{equation}
The subscript on the right-hand-side is to indicate that the $\delta$-function
is smeared over a time interval
$1/\Delta^{\Omega}_{n(t_m)}$.
Inserting eq. (\ref{ndist}) into eq. (\ref{rateeq}) we obtain
\begin{eqnarray}
\ddot{\phi}({\bf k}_n,t) &+&
(3H + \Gamma) {\dot \phi} ({\bf k}_n,t)
+(e^{-2Ht}{\bf k}_n^2+m^2 - \frac{2 \Gamma \Delta_{\mu}}{\pi}
+\frac{3\Gamma H}{2})\phi({\bf k}_n,t)
\nonumber\\
&+&\frac{\delta V_F}{\delta \phi(-{\bf k}_n,t)}
+2\Gamma e^{-\frac{3}{2} H (t-t_m)} \phi({\bf k}_n,t_m)
\delta_{1/\Delta^{\Omega}_{n(t_m)}}
(t-t_m) = \eta ' ({\bf k}_n,t; t_m)
\label{phieqlim}
\end{eqnarray}
where
\begin{equation}
\eta ' ({\bf k}_n,t;t_m) \equiv -\int d \mu N(\mu) \alpha (\mu)
\eta^0 ({\bf k}_n,t;t_m; \mu).
\end{equation}
In the limit of zero expansion, $H=0$, we obtain from
eq. (\ref{phieqlim})
\begin{eqnarray}
\ddot{\phi}({\bf k}_n,t) &+&
\Gamma {\dot \phi} ({\bf k}_n,t)
+({\bf k}_n^2+m^2 - \frac{2 \Gamma \Delta_{\mu}}{\pi})
\phi({\bf k}_n,t)
\nonumber\\
&+&\frac{\delta V_F}{\delta \phi(-{\bf k}_n,t)}
+2\Gamma \phi({\bf k}_n,t_m)
\delta_{1/\Delta^{\Omega}_{n(t_m)}}
(t-t_m) = \eta^{'}_0({\bf k}_n,t)
\label{phieqlim2}
\end{eqnarray}
where $\eta'_0({\bf k}_n,t)$ is as in eq. ({\ref{phieqlim}) but
with $H=0$ in
eq. (\ref{etafree}).
Aside from the last term on the left-hand-side, which is
nonzero only for a very short time interval,$1/\Delta_{\mu}$,
this is the standard form of the Langevin equation
and the type we used in \cite{wi}.
In the limit $T \rightarrow \infty $ we  find from
eq. (\ref{symmcor}) and eq. (\ref{ndist})
\begin{equation}
\frac{1}{2} \langle \langle
\eta_0({\bf k}_n,t) \eta_0(-{\bf k}_n,t')
+ \eta_0(-{\bf k}_n,t') \eta_0({\bf k}_n,t)
\rangle \rangle_{T\rightarrow \infty}
\rightarrow 2 L^3 \Gamma T \delta(t-t'),
\label{fdvalue}
\end{equation}
which verifies the fluctuation-dissipation theorem \cite{cw}.

\section{Interpretation}
\label{sec.inter}

In the last section we obtained the generalized
Langevin equation (\ref{rateeq}),
which induces fluctuations on the inflaton as well as frictionally
damps its motion.  To obtain this equation, we had to introduce a
heat reservoir of light particles that interacted with the inflaton.
In \cite{wi} one form of rate equation (\ref{rateeq}),
namely eq. (\ref{phieqlim2}), was shown to be numerically successful
for inflaton dynamics.  This evidence provides sufficient motivation
to seek an interpretation of the heat reservoir in the context of
particle physics.  The heat reservoir of the last section
is a toy model.  Here we want to think generally about a second
system of light particles which acts as a random force on
the inflaton.

For the dynamics described in the last section to be realizable
in the real world, in particular two properties
are needed for the reservoir.  Firstly there must be some mechanism
available during inflation to produce the reservoir particles.
Secondly these particles must interact rapidly on the scale
of the expansion time $1/H$.
In section (III) we assumed that they could thermalize.
This stringent of a condition is not needed
for eq. (\ref{rateeq}) to describe stochastic evolution
of the inflaton.  Nevertheless, we keep with
our thermalization assumption and will see how close we can come.
In addition to these two properties, a less important
point of interest is what direct interpretation the explicit reservoir
Hamiltonian (\ref{resli}) in our simple model has.

To address these issues, we first make
three general observations.  Firstly,
the zero mode of the inflaton is evolving quasi-statically
during its roll down the potential. Secondly, quantum processes
will occur within micro-physical scales, so for
physical distances and cosmic time less than $1/H$ at any e-fold of inflation.
The third observation is based on our analysis in \cite{wi}.
There we found that the energy density of the heat reservoir was close to the
GUT scale $M_{\rm GUT}^4$.  In particular the temperature
range of the thermal scenarios we considered that were consistent
with observation was
$T \sim (0.01 - 0.03) M_{\rm GUT}$. This implies
for a single light species a corresponding energy density in
the range
\begin{equation}
\rho_r \sim (10^{-8} - 10^{-6}) M^4_{\rm GUT}.
\label{rocklim}
\end{equation}
For $M_{\rm GUT}=10^{15}$ GeV= $1.782 \times 10^{-9}$ g and for the
lower bound on $\rho_r$ this means
\begin{eqnarray}
\rho_r \geq 10^{-8} M^4_{\rm GUT} &=& 10^{52} \ {\rm GeV}^4 =
10^{66} {\rm g/cm}^3
\nonumber \\
&\approx& 10^{51} \rho_{{\rm neutron \ star}} \approx
10^{50} \rho_{{\rm HI-collision}},
\label{enres}
\end{eqnarray}
where $\rho_{{\rm neutron \ star}}$ and
$\rho_{{\rm HI-collision}}$ are the energy densities for
respectively a neutron star and an upper bound estimate
for a heavy-ion collision.
For our present purpose, we make the general observation that the
energy density of the heat reservoir is large on the scale of
the largest probed energy densities.

Observation one provides suitable conditions for particle
creation from quantum decay of the inflaton similar to pair-production
in a strong quasi-static electric field.  This gives
a mechanism for producing the particles in the heat reservoir.
An outcome is
that the description of the reservoir's
state must be inherently statistical.

Having made a particle association for the heat reservoir, for completeness
of our model in section (III), we make an interpretation
of $H_R$ in eq. (\ref{hres}).
The most direct statement is that
$H_R$,
with an appropriate spectral weight
function eq. (\ref{spec}), is a general representation of scalar
quasiparticle excitations in the noninteracting limit.
On the other hand,
thinking in terms of elementary fields,
if we apply dimensional counting to $H_I$ in eq. (\ref{resli})
we find that the reservoir fields $\eta_i({\bf x},t)$ carry
engineering dimension 3.  This suggests that in
our simple model each reservoir field
represents a fermion-antifermion pair (ie. $q {\bar q}$, $e {\bar e}$
etc...), which interacts with the inflaton through a Yukawa coupling.
$H_R$ then has the interpretation of an effective
free field Hamiltonian for paired $f {\bar f}$ states.
In QCD these would be the simplest color singlet states
constructible from quarks.
Beyond these "kinematic" statements, we can not say more at present.
Almost nothing is known about matter
at the energy densities of eq. (\ref{enres}).  We therefore must leave
unanswered to what extent $H_R$ in eq. (\ref{hres}) and the spectral
weight function in eq. (\ref{specconst}) approximate a heat reservoir system
which can be derived from a realistic particle physics Lagrangian.

Returning from this digression, next we will examine what expectation
there is for thermalization in the reservoir.
For this let us review the situation up to now.
Eq. (\ref{rateeq}) contains the dynamics
that governs energy transfer between the
reservoir and the inflaton.  The full internal dynamics of the inflaton
system in isolation are also treated.  Thus
eq. (\ref{rateeq}) is valid for an
arbitrary scalar inflaton potential and for reservoir fields
that do not interact directly with each other.  We discussed
in section (III) that in the rapid expansion environment during inflation
the internal interactions of an actual reservoir system are
essential for distributing any influx of energy.
Having established above that the state of the reservoir is inherently
statistical, knowledge of internal reservoir dynamics is
in particular important
for addressing the question of thermalization. In the last section we got
around
the problem by assuming that thermalization occurs and then made
suitable adjustments to our equations.

To check the assumption
of thermalization is a difficult problem
even if one is given more dynamical details
about the reservoir.  However based on observations two and three
we can get oriented with the scales involved in the problem.
For this let us consider an unjustifiable classical model
of light hard rocks of average energy $E$, center-of-mass two-body
elastic cross section $\sigma$,
and at a energy density $\rho_{\rm rocks}$ of order eq. (\ref{enres}).
The mean free path in this system is
\begin{equation}
l_{\rm rocks} = \frac{E}{\rho_{\rm rocks} \sigma},
\end{equation}
which for relativistic rocks implies a collision frequency
$\omega_{\rm rocks} = 1/l_{\rm rocks}$.

We assume that the elastic cross section is independent of
energy.  Up to logarithmic corrections, this corresponds
to Regge asymptotic behavior from single pomeron exchange with
slope one (for a review please see \cite{agg} and references
therein).  For a conservative lower bound to elastic
$pp$-scattering at high energy,
we use $\sigma_{el} = 10 ~{\rm GeV}^{-2}$.
The energy of a typical light particle at temperature $T$
is $E \sim T$.  Taking  the lower bound from eq. (\ref{enres})
$\rho_{\rm rocks} = 10^{52} ~{\rm GeV}^{4}$ which
corresponds to  $T=0.01M_{\rm GUT}$,
we find
\begin{equation}
\omega_{\rm rocks} = 10^{40} ~{\rm GeV} \approx 10^{30} H.
\label{omrocks}
\end{equation}
It is known that most of the elastic cross section $\sigma_{el}$
is increasingly in the forward direction with increasing
energy.   Accounting for this,
we take the average momentum transfer to be
$|t| = 0.025 {\rm GeV}^2$.  Let us assume statistical independence
between collisions for a typical light particle.
Then within one e-fold,
the root-mean-square energy that the particle transfers will be about
an order of magnitude greater than its total energy and
it samples all angular directions in phase space.
Optimistically the range for eq. (\ref{omrocks})
could be increased by a factor $10^3$ by using the upper
bound on $\rho_{\rm rocks}$ in eq. (\ref{rocklim})
and increasing $\sigma_{el}$ by over a factor $10$.
This range presents a case that some sort of
GUT-plasma could thermalize within an e-fold at
GUT-scaled energy densities.

\section{Mechanical Analogue}
\label{sec.meanl}

In section (III) we derived a quantum Langevin equation, but we can think
of it as a classical Langevin equation.  For this we can make a picture
of the dynamics for a classical inflaton field $\phi^c({\bf x},t)$.
We define the order parameter in the classical system as
\begin{equation}
\phi^c(t) \equiv \int_V \frac{d^3{\bf x}}{V}\phi^c({\bf x},t),
\end{equation}
and it will sometimes be denoted as $\phi^c$.
For the classical theory we will make a mechanical model and then associate it
with the classical field.  We will work  in the infinite volume limit
where correspondence between the classical and quantum fields
can be stated more simply.

Let us first recall some details about the inflaton Hamiltonian.
In thermal scenarios the symmetry breaking potential $V(\phi)$ can be
of the standard double-well type without an ultra-flat region
as demonstrated in \cite{wi}.
The relevant magnitudes which define the potential are the
same as in the standard scenario.  Thus the vacuum energy
$V(0) \sim M^4_{\rm GUT}$, the minima of the wells are at
$\phi_m \approx \pm M_{\rm GUT}$, and the curvature at
the minima are $V''(\phi_m) \approx M^2_{\rm GUT}$.
Inflation begins when the inflaton tunnels out of the meta-stable minima
at $\phi=0$ to $0 < \phi_i < \phi_m$.  We will define
this moment as the origin of cosmic time $t$.  Our classical
analogy begins at t=0.

With these preliminary remarks, let us state our classical analog model.
We picture the classical scalar field as composed of heavy particles
of mass $M_{\phi} = \sqrt{V''(\phi_m)}$.  They sit in a cubic
lattice arrangement with massless  (anharmonic if
preferred) springs connecting them.
Kinetic energy is associated with the particle masses and
their motional energy.  Potential energy is
associated with the energy in the massless springs.
The order parameter $\phi^c(t)$
is interpreted to be related to the number density
$n_{\phi}(t)$ of $\phi$-particles.  To obtain a precise relationship
we will equate the order parameter with the vacuum expectation
value of the quantum Heisenberg scalar field at t=0 as
\begin{equation}
\phi^c(0)= \lim_{V \rightarrow \infty}
\langle 0| \int_V \frac{d^3 {\bf x}}{V} \phi({\bf x},0)
|0 \rangle.
\end{equation}
We will assume that the vacuum is a coherent state of zero momentum
particles.  Using the number density operator
\begin{equation}
{\hat n} \equiv \lim_{V \rightarrow \infty} \frac{{\hat N}}{V}
= \langle \rangle - \lim_{V \rightarrow \infty}
\frac{1}{V} \sum_n a^{\dagger}_n a_n,
\label{nlim}
\end{equation}
we arrive at
\begin{equation}
\phi^c(0) = \sqrt{\frac{2n_{\phi}(0)}{M_{\phi}}}.
\end{equation}
In eq. (\ref{nlim}) $\langle \rangle - {\rm lim}$
means evaluate the operator matrix element in finite
volume and then take the limit $V \rightarrow \infty$.
In our classical model we define the order parameter
at all time with this relation,
\begin{equation}
\phi^c(t) \equiv \sqrt{\frac{2n_{\phi}(t)}{M_{\phi}}}.
\label{numden}
\end{equation}
If one prefers, this can be regarded as an unmotivated definition,
which relates the the classical mechanical model with our
classical field model.

The initial conditions at $t=0$ are the order parameter is at
$\phi^c(0) = \phi_i$, the masses are at rest with number density
given by eq. (\ref{numden}), and the springs are elongated such that the
cubic symmetry is maintained but the fundamental cell is bigger.
When the system is released a breathing mode commences, in which
the fundamental cell contracts to a minima and the expands back
to its original size at $t=0$ and so on.
Correspondingly, the number density oscillates.

To this system, we will add many light $\eta$-particles of
mass $m_{\eta} \ll M_{\phi}$ that are thermalized
at temperature T.  This system forms the heat reservoir.
The $\eta$-particles randomly will impart momentum
on the $\phi$-particles.  This will
perturb the breathing mode of the $\phi$-particles
as well as excite transverse modes on the $\phi$-lattice.
The system so described is similar to that for Brownian
motion, except now there are several Brownian particles
and they also have an interaction amongst themselves.
With minor changes, one can understand the rest of the statistical mechanics
for our system from Brownian motion.
Slow-roll motion in our classical model corresponds to
over-damping, in which the fundamental cell slowly
contracts to equilibrium with no
oscillations.  In the process the potential
energy in the $\phi$-system is transferred to the heat
reservoir.
One can also expand this picture to include inflationary
expansion and $\eta$-particle creation, but we will
leave that for the reader's thoughts.

\section{Conclusion}
\label{sec.conc}

In section (III) we obtained the rate eq. (\ref{rateeq}) which describes the
dynamics of the inflaton's comoving modes while they are
sub-horizon size.  For the zero mode, eq. (\ref{rateeq}) is therefore
unjustified, since it does not treat super-horizon scale physics.
What one needs is an equation that describes the transfer of potential
energy from the inflaton's vacuum during its roll down to the potential
minima.  In our model, the energy is transfered not only
into the inflaton's kinetic energy, but also into the fields
of the reservoir.

In a local patch and for a short time interval, eqs. (\ref{rateeq})
would be valid for all modes which are ad-hocally defined with respect
to the patch.  A network of such patches could be constructed
to cover the inflating universe for some chosen cosmic time interval.
In each patch the respective Langevin equation could be solved
with solutions matched at the boundaries.
To compute the initial sub-horizon scale energy density perturbations,
recall that they are determined by the sub-horizon scale modes.
Therefore this computation could be done from the solution
of the Langevin equation from any one patch.
For the zero mode, there are two points of interest.
First we want to know the global energy
loss by the inflaton.  Due to the lack of coherence
between patches, the expectation value of
the inflaton (order parameter) need not be the same in different
patches.
Different local regions
of the universe could then be in different stages of inflation,
depending on the local order parameter.
For computing the total change in vacuum energy, one could
sum the result from the independent patches.  However
turning to the second point, for inflation to solve the smoothness
problem, one requires the order parameter to be approximately
globally homogeneous.
One may argue that the local conditions in all
patches would be similar, so that the order parameter should not
differ by too much in different patches.  To verify
this would require
further investigation within specific models.
In particular, to meet observational
constraints, one wants to know the degree of homogeneity of
the order parameter within an inflating region for
a duration of at least 70 e-folds.
All statements so far apply equally well to scenarios based
on both quantum and thermal fluctuations.
There is one difference.
Thermal scenarios,
do not require the quantum state of the inflaton's
zero mode to be coherent over the entire inflationary
patch. This is required in the standard scenario for reheating.

The derivation in section (III) is formally valid for a fundamental
or effective inflaton Lagrangian $L_S$ in eq. (\ref{langs}).
Given an initial fundamental Lagrangian for the inflaton, one
anticipates that to formulate a consistent
stochastic process, it will require high frequency interactions
of the inflaton with itself and with other fields to be
treated first and appear in the form of an effective Lagrangian
$L_S$ in eq. (\ref{langs}).  We have not developed in this
paper any methods for obtaining the appropriate effective Lagrangian
one should use for $L_S$.

In \cite{wi} we assumed near equilibrium
dynamics applied everywhere.  We therefore used the finite temperature
effective potential in $L_S$.  However nonequilibrium considerations
such as done in \cite{gleiser} for flat space may be important to
extend for the inflationary era of expanding space.
Generally for a scalar inflaton field coupled to gauge fields,
such as in GUT theories, one should expect quantum corrections
from self-coupling and quantum plus thermal corrections
from coupling to the gauge fields to give a modified potential
in $L_S$.  Since the relevant temperature scale for slow-roll
dynamics is of order or less than the inflaton mass,
one does not expect thermal corrections from the inflaton's
self-coupling to be substantial.

To summarize, in this paper we showed that the presence of a sizable thermal
component during inflation is energetically allowed.  Following
our hypothesis in \cite{bf1}, we treated the thermal  component
as a heat reservoir, which interacted with the inflaton
in the manner of Langevin dynamics.  We demonstrated in
section (III) that the Langevin equation postulated in \cite{bf1}
could be derived from quantum field theory.
In \cite{wi}
we showed that such a dynamical arrangement,
for the specific case of a double-well scalar potential,
was numerically successful in describing observational data
without fine tuning the coupling constant.

One of the improvements suggested by our calculations there
was to have a time dependent temperature that smoothly varied
by a factor 2-3 over the course of the roll-down period.
Not only was this numerically preferable, but it seems
that thermodynamics does not restrict
some variation in the temperature just before to just after
inflation.  As such the general dynamical equations during inflation
should be able to accommodate such a variation, although
specific models may make specific predictions.

In \cite{wi} the reservoir was defined by its thermodynamic
properties, but was not given a dynamical representation.
Thus we had no underlying mechanism that could explain temperature
variations.  In section (III) our model supplies an
explanation.   However it is clear from our treatment there,
that it was an arbitrary decision to hold the reservoir's
energy density constant with respect to physical volume
all throughout inflation.  Just as well
within the same dynamical framework,
it could increase or decrease and in different time
periods do either.  However this is as far as the model in its present form
can go.  On the plus side, the model goes far enough to
dynamically explain a time dependent
reservoir energy density.  However the model is not
sufficiently constrained to do better than this.
The options to improve this situation are to determine
a further constraint from either formal means or phenomenological
motivation.
\section*{Acknowledgments}

I thank J. Pullin, L. Z. Fang and A. Ashtekar for helpful discussions.
This work was supported by the U.S. Department
of energy under grant numbers DE-FG02-90ER-40577 and
DE-FG02-93ER40771.

\appendix
\section{Statistical mechanics of free fields}

Some statistical mechanical properties of free field, which
are often used in the text, are given here. We consider
a free field Hamiltonian defined in a box of volume $V_L$ centered
at the origin with sides at $\pm L/2$ in all three spatial directions.
We give results in the discrete form in $V_L$ for finite but
large $L$ and then in the infinite volume limit, $V_{\infty}$,
where $L \rightarrow \infty$.  The free field Hamiltonian
which we study is
\begin{equation}
H=\int_{V_L} d^3{\bf x} \frac{1}{2} \left[ \pi_{\chi}^2({\bf x},t)+
({\bf \nabla} \chi ({\bf x},t))^2
+m^2\chi^2({\bf x},t) \right] + C,
\label{haml}
\end{equation}
where
C is a constant that later will be set so that the lowest energy
state is at zero.  The fields satisfy the equal time canonical
commutation relations (CCR)
\begin{eqnarray}
[\chi ({\bf x},t), \chi ({\bf x}',t)]&=&0
\nonumber\\
{[}\pi_{\chi} ({\bf x},t), \pi_{\chi} ({\bf x}',t)]&=&0
\nonumber\\
{[}\pi_{\chi} ({\bf x},t), \chi ({\bf x}',t)]&=&-i\delta^{(3)}({\bf x}-{\bf
x}').
\label{ccr}
\end{eqnarray}
The field have the expansion in terms of creation and
annihilation operators
\begin{eqnarray}
\chi({\bf x},t)&=&\frac{1}{L^{3/2}}\sum_{n_x,n_y,n_z=-\infty}^{\infty}
\frac{1}{\sqrt{2\omega_n}}[a_ne^{i({\bf k}_n \cdot {\bf x} -\omega_nt)}
+a^{\dagger}_ne^{-i({\bf k}_n \cdot {\bf x} -\omega_nt)}]
\nonumber\\
\pi_{\chi}(x,t)&=&\frac{-i}{ L^{3/2}}\sum_{n_x,n_y,n_z=-\infty}^{\infty}
\sqrt{\frac{\omega_n}{2}}[a_ne^{i({\bf k}_n \cdot {\bf x} -\omega_nt)}
-a^{\dagger}_ne^{-i({\bf k}_n \cdot {\bf x} -\omega_nt)}]
\label{chiexp}
\end{eqnarray}
where
\begin{equation}
{\bf k}_n \equiv \frac{2\pi {\bf n}}{L},
\end{equation}
\begin{equation}
{\bf n} \equiv (n_x,n_y,n_z),
\end{equation}
\begin{equation}
\omega_n \equiv \sqrt{{\bf k}_n^2+m^2},
\end{equation}
and from eq. (\ref{ccr})
\begin{equation}
[a_n,a_{n'}]=[a^{\dagger}_n,a^{\dagger}_{n'}]=0
\end{equation}
\begin{equation}
[a_n,a^{\dagger}_{n'}]=\delta_{nn'}.
\label{creann}
\end{equation}

We define the Fourier transform of the fields as
\begin{equation}
\chi ({\bf k}_n,t) \equiv  \int_{V_L} d^3{\bf x} \chi ({\bf x},t)
e^{-i{\bf k}_n \cdot {\bf x}}
\label{fourchi}
\end{equation}
\begin{equation}
\pi_{\chi} ({\bf k}_n,t) \equiv \frac{1}{ L^3} \int_{V_L} d^3{\bf x}
\pi ({\bf x},t)
e^{-i{\bf k}_n \cdot {\bf x}}.
\label{fourpi}
\end{equation}
{}From eq. (\ref{ccr}) one can verify that
the Fourier space modes of the fields satisfy the equal time CCR
\begin{eqnarray}
[ \chi({\bf k}_n,t), \chi({\bf k}_{n'},t) ] &=&0
\nonumber\\
{[} \pi_{\chi}({\bf k}_n,t), \pi_{\chi}({\bf k}_{n'},t) ]&=&0
\nonumber\\
{[} \pi_{\chi}({\bf k}_n,t), \chi(-{\bf k}_{n'},t) ] &=& -i\delta_{nn'}.
\label{fccr}
\end{eqnarray}
The expansion of the Fourier-modes in terms of the
creation-annihilation operators
is
\begin{eqnarray}
\chi({\bf k}_n,t)&=&\frac{L^{3/2}}{\sqrt{2\omega_n}}
[a_ne^{-i\omega_n t}
+a^{\dagger}_{-n}e^{i\omega_nt}]
\nonumber\\
\pi_{\chi}({\bf k}_n,t)&=&\frac{-i}{L^{3/2}}\sqrt{\frac{\omega_n}{2}}
[a_ne^{-i\omega_nt}
-a^{\dagger}_{-n}e^{i\omega_nt}].
\label{kmodes}
\end{eqnarray}

The constant C in eq. (\ref{haml}) is be defined as
\begin{equation}
C=-\frac{1}{2}\sum_n \omega_n.
\end{equation}
Two equivalent forms of the Hamiltonian H in eq. (\ref{haml}) are
given below.  Writing
\begin{equation}
H=\sum_n H_n =\sum_{n} H_{k_n},
\end{equation}
where in terms of the creation-annihilation operators
in eq. (\ref{creann}),
\begin{equation}
H_n= \omega_n a^{\dagger}_n a_n
\label{ham1}
\end{equation}
and in terms of the Fourier-modes in eq. (\ref{kmodes})
\begin{equation}
H_{k_n}=\frac{L^3}{2} \pi_{\chi}({\bf k}_n,t)\pi_{\chi}(-{\bf k}_n,t)
+ \frac{1}{2 L^3}
({\bf k}_n^2+m^2)\chi({\bf k}_n,t)\chi(-{\bf k}_n,t).
\label{ham2}
\end{equation}

The partition function, $Z \equiv {\rm Tr} \left(e^{-H/T}\right)$, evaluated
in the creation-annihilation
basis is a product $Z=\prod_n Z_n$, where
\begin{equation}
Z_n = {\rm Tr} \left(e^{-H_n/T}\right)=\frac{1}{1-e^{{-\omega_n}/{T}}}.
\end{equation}

The two-point thermal correlation function for the creation-annihilation
operators is
\begin{eqnarray}
\frac{1}{Z_n}{\rm Tr}
\left(e^{\frac{-H_n}{T}}a^{\dagger}_n a_{n'}\right) & \equiv &
\langle \langle a^{\dagger}_n a_{n'} \rangle \rangle_T
= \delta_{nn'} \frac{1}{e^{{\omega_n}/{T}}-1}
\nonumber\\
\langle \langle a_n a_{n'}\rangle \rangle_T
&=& \langle \langle a^{\dagger}_na^{\dagger}_{n'}\rangle \rangle_T=0
\end{eqnarray}
and for the Fourier-modes is
\begin{eqnarray}
\frac{1}{Z} {\rm Tr}
\left(e^{\frac{-H}{T}}\chi({\bf k}_n,t)\chi({\bf k}'_n,t')\right) &\equiv&
\langle \langle \chi({\bf k}_n,t)\chi({\bf k}'_n,t'\rangle \rangle_T
\nonumber\\
&=&
\delta_{n,-n'} \frac{L^3}{2\omega_n}
\left[\coth\left(\frac{\omega_n}{2T}\right)\cos \omega_n(t-t')
-i\sin \omega_n(t-t') \right]
\nonumber\\
\langle \langle \pi_{\chi}({\bf k}_n,t) \pi_{\chi}({\bf k}'_n,t')
\rangle \rangle_T &=&
\delta_{n,-n'} \frac{\omega_n}{2 L^3}
\left[\coth \left(\frac{\omega_n}{2T}\right)\cos \omega_n(t-t')
-i\sin \omega_n(t-t') \right]
\nonumber\\
\langle \langle \chi({\bf k}_n,t) \pi_{\chi}({\bf k}'_n,t')
\rangle \rangle_T &=&
\delta_{n,-n'} \frac{1}{2} \left[
\coth\left(\frac{\omega_n}{2T}\right)\sin \omega_n(t-t')
+i\cos \omega_n(t-t') \right].
\label{chidist}
\end{eqnarray}
{}From the last relation note that
\begin{equation}
\langle \langle [ \pi_{\chi}({\bf k}_n,t), \chi({\bf k}'_n,t') ]
\rangle \rangle_T =
-i\delta_{n,-n'} \cos \omega_n(t-t').
\end{equation}

The total ensemble averaged energy of the thermalized system is
\begin{equation}
U_{V_L}(T)= \sum_n U^n_{V_L}(T)
\label{uvl}
\end{equation}
where
\begin{equation}
U^n_{V_L}(T) = \omega_n \langle \langle a^{\dagger}_n a_n \rangle \rangle_T
= \frac{\omega_n}{e^{\omega_n/T}-1}.
\label{unvl}
\end{equation}

Let us examine the infinite volume limit.  The creation and annihilation
operators must be rescaled as
\begin{equation}
c({\bf k}_n) \equiv \left(\frac{L}{2 \pi}\right)^{3/2} a_n
\label{cann}
\end{equation}
and similarly for $a^{\dagger}_n$.
Substituting eq. (\ref{cann}) into eq. (\ref{chiexp})
and identifying
\begin{equation}
\left(\frac{2\pi}{L} \right)^3 =
\left(\frac{2 \pi \Delta n}{L}\right)^3_{L \rightarrow \infty}
\rightarrow d^3{\bf k},
\end{equation}
we get the continuum form
\begin{equation}
\chi({\bf x},t)=\int \frac{d^3{\bf k}}{[(2 \pi)^3 2 \omega_{\bf k}]^{1/2}}
[c({\bf k}) e^{i({\bf k}_n \cdot x -\omega_{\bf k}t)}
+c^{\dagger}({\bf k}) e^{-i({\bf k}_n \cdot x -\omega_{\bf k}) t)}]
\end{equation}
and similarly for $\pi_{\chi}({\bf x},t)$.
For the momentum space Fourier transforms of the fields in the infinite volume
limit, $V_L$ must be replaced by $V_{\infty}$ in eqs. (\ref{fourchi})
and (\ref{fourpi})
and $\pi_{\chi}({\bf k},t)$ must be defined without the factor
$1/L^3$.  The Hamiltonian in the forms eq. (\ref{ham1}) and eq. (\ref{ham2})
in $V_{\infty}$ are respectively
\begin{eqnarray}
H &=&  \int d^3 {\bf k} \ \omega_{\bf k} c^{\dagger}({\bf k}) c({\bf k})
\nonumber\\
&=&\int \frac{d^3{\bf k}}{(2 \pi )^3} \frac{1}{2}
\left[ \pi_{\chi}({\bf k},t)\pi_{\chi}(-{\bf k},t)+
({\bf k}^2+m^2)\chi({\bf k},t)\chi(-{\bf k},t) \right].
\end{eqnarray}
The energy density in $V_{\infty}$ is obtained as
\begin{equation}
u(T) = \lim_{L \rightarrow \infty} \frac{U_{V_L}(T)}{L^3}.
\end{equation}
{}From eq. (\ref{uvl}) and (\ref{unvl}) we find
\begin{equation}
u(T) = \int \frac{d^3{\bf k}}{(2 \pi)^3}
\frac{\omega_{\bf k}}{e^{\omega_{\bf k}/T}-1}.
\end{equation}
The familiar formula for the energy density of electromagnetic
radiation is twice this, due to two polarization states for the photon,
and with $m=0$ so that $\omega_{\bf k} = |{\bf k}|$.

\section{Alternative Derivation of the Rate Equation}

For a system with a nearly harmonic potential,
such as within one well of a double-well potential,
one can derive the Langevin
equation with a general reservoir Hamiltonian
$H_R$ in the regime
\begin{equation}
H \ll \Gamma \ll \omega,
\label{gamlim}
\end{equation}
where $H$ is the Hubble constant, $\Gamma$ is the dissipative constant
and $\omega$ is the frequency of the system with no anharmonic terms in
the potential.
We perform the derivation in this appendix. It follows
the method of \cite{sen}.  As in section III,
it is done in a cube for the three spatial directions,
which is centered at the origin with sides at $\pm L/2$
in each direction.

The Hamiltonian we examine is
\begin{equation}
H_T=H_S + H_R + H_I,
\label{totham}
\end{equation}
where $H_S$ is the system Hamiltonian given for the inflaton
field in eq. (\ref{hsys}), $H_R$ is
the Hamiltonian for the reservoir system when it is
uncoupled to the scalar field system, and
\begin{equation}
H_I=
\alpha \int_V d^3{\bf x} e^{3Ht} \Omega({\bf x},t)\phi({\bf x},t)
\label{hi2}
\end{equation}
where $\Omega({\bf x},t)$ is a Hermitian operator
made of any field or composite fields of the reservoir
system that couples to the scalar field.
For our derivation,
we do not need specific details about $H_R$.  It may
well represent a strongly interacting system with respect to
$\Omega({\bf x},t)$.  Expressed in momentum space, eq. (\ref{hi2})
is
\begin{equation}
H_I=\frac{e^{3Ht}}{L^3} \sum_n \alpha
\Omega({\bf k}_n,t) \phi(-{\bf k}_n,t).
\end{equation}

Postulating the standard equal time CCR of eqs. (\ref{ccr})
and (\ref{fccr}),
the equations of motion for the inflaton are
\begin{equation}
{\dot \phi}({\bf k}_n, t) = i[H_T,\phi({\bf k}_n,t)]=
e^{-3Ht}L^3 \pi_{\phi}({\bf k}_n,t)
\label{firstphi}
\end{equation}
\begin{equation}
{\dot \pi}_{\phi} ({\bf k}_n, t) = i[H,\pi_{\phi}({\bf k}_n,t)]=
-\frac{e^{3Ht}}{L^3}[\omega_n^2(t) \phi({\bf k}_n,t)
+\frac{\delta V_F}{\delta \phi(-{\bf k}_n,t)}
+\alpha \Omega({\bf k}_n,t) ]
\label{firstpi}
\end{equation}
where
\begin{equation}
\omega_n(t) =\sqrt{ e^{-2Ht}{\bf k}_n^2+m^2}.
\end{equation}
This implies the second order equation of motion
\begin{equation}
\ddot{\phi} ({\bf k}_n, t) +3H{\dot \phi}({\bf k}_n,t)
+\omega_n^2(t) \phi({\bf k}_n,t)+
\frac{\delta V_F}{\delta \phi(-{\bf k}_n,t)}
+ \alpha \Omega ({\bf k}_n,t) = 0.
\label{phi2eq}
\end{equation}

We derive the effective equation of motion
for a given comoving mode of the inflaton.
In the parametric regime eq. (\ref{gamlim}),
the time interval $1/\Gamma \ll \Delta t \ll 1/H$
is sufficient to study the relaxational
dynamics of the inflaton.  Thus we can drop the second term in eq.
(\ref{phi2eq}) above and assume that the comoving wavenumber
${\bf k}_n$ is constant
at $e^{-2Ht}{\bf k}_n = {\bf k}_n^m$, where ${\bf k}_n^m$
is the initial value in the time interval m of interest.
Dropping the superscript $m$,
the resulting second order equation of motion is
\begin{equation}
\ddot{\phi} ({\bf k}_n, t)
+\omega_n^2\phi({\bf k}_n,t)+
\frac{\delta V_F}{\delta \phi(-{\bf k}_n,t)}
+ \alpha \Omega ({\bf k}_n,t) = 0,
\label{phi2eqres}
\end{equation}
where
\begin{equation}
\omega_n = \sqrt{{\bf k}_n^2+m^2}.
\end{equation}
This can be obtained from the first order equations in
eqs. (\ref{firstphi}) and (\ref{firstpi}) by the replacement
\begin{equation}
e^{-2Ht} {\bf k}_n^2 \rightarrow
e^{-2Ht_m} {\bf k}_n^2
\label{shift}
\end{equation}
in $\omega_n(t)$ in eq. (\ref{firstpi}) and by
setting $H=0$ everywhere else.

Doing this, the solution to
eqs. (\ref{firstphi}) and (\ref{firstpi}) are
respectively
\begin{eqnarray}
\phi({\bf k}_n,t) &=& \phi_0({\bf k}_n,t) - \frac{\alpha}{\omega_n}
\int_0^t dt' \Omega({\bf k}_n,t') \sin \omega_n(t-t')
\nonumber\\
&-&\frac{1}{\omega_n}\int_0^t dt'
\frac{\delta V(\phi)}{\delta \phi(-{\bf k}_n,t')} \sin \omega_n (t-t')
\label{A1}
\end{eqnarray}
\begin{eqnarray}
\pi({\bf k}_n,t) &=& \pi_0({\bf k}_n,t) - \frac{\alpha}{\omega_n}
\int_0^t dt' \Omega({\bf k}_n,t') \cos \omega_n(t-t')
\nonumber\\
&-&\frac{1}{\omega_n}\int_0^t dt'
\frac{\delta V(\phi)}{\delta \phi(-{\bf k}_n,t')} \cos \omega_n (t-t'),
\label{A2}
\end{eqnarray}
where $\phi_0$ and $\pi_0$ are the solutions of eqs. (\ref{firstphi})
and (\ref{firstpi}) with $V_F=0$ and $\alpha=0$. Again for
notational convenience we denote the physical momentum simply
by ${\bf k}_n$.  The associated comoving mode can be computed
at time $t_m$ from eq. (\ref{shift}).

The inflaton (system) can be treated as a small perturbation on
$\Omega({\bf k}_n,t)$, however full account must be taken
of the heat reservoir's effect on the inflaton.
Thus we will account for the lowest order effect of the system on
the reservoir coordinate $\Omega({\bf k}_n,t)$, which gives
\begin{eqnarray}
\Omega ({\bf k}_n,t) &=& e^{-iHt} \Omega ({\bf k}_n,0) e^{iHt}
\nonumber\\
&=& \Omega_0({\bf k}_n,t) + \frac{i \alpha}{L^3}
\left\{ \Omega_0({\bf k}_n,t) \sum_{n'} \int^t_0 dt' \phi({\bf k}_{n'},t')
\Omega_0(-{\bf k}_{n'},t') \right.
\nonumber\\
&-& \left. \sum_{n'} \int^t_0 dt' \phi({\bf k}_{n'},t-t')
\Omega_0(-{\bf k}_n,t-t') \Omega_0({\bf k}_n,t) \right\}
\end{eqnarray}
where
\begin{equation}
\Omega_0({\bf k}_n,t) = e^{-iH_R t} \Omega ({\bf k}_n,0) e^{iH_R t}
\end{equation}
is the time development of $\Omega$ in the uncoupled reservoir.

Substituting this into the second term on the right-hand-side of
eq. (\ref{A1}), this term becomes
\begin{equation}
\frac{\alpha}{\omega_{n}} \int^t_0 dt' \Omega({\bf k}_n,t')
\sin \omega_n(t-t') =
\frac{\alpha}{\omega_n} \int^t_0 dt' \Omega_0({\bf k}_n,t')
\sin \omega_n(t-t') + L({\bf k}_n,t),
\label{2term}
\end{equation}
where
\begin{eqnarray}
L({\bf k}_n,t)= \frac{i\alpha^2}{\omega_n L^3} \sum_{n'}\int^t_0 dt'
\sin \omega_n (t-t')
\int^{t'}_0 dt''& & \left[\phi({\bf k}_{n'},t'') \Omega_0({\bf k}_n,t')
\Omega_0(-{\bf k}_{n'},t'') \right.
\nonumber \\
&-& \left. \phi({\bf k}_{n'},t'-t'') \Omega_0(-{\bf k}_{n'},t'-t'')
\Omega_0({\bf k}_{n},t') \right].
\end{eqnarray}

We now take the statistical expectation value of
eq. (\ref{phi2eqres})
with respect to the free states of the heat reservoir, which
are assumed to obey a canonical distribution.
The matrix elements of $\Omega_0({\bf k}_n,t)$ between
state $<a|$ and $|b>$ of the uncoupled reservoir system are
written as
\begin{equation}
<a| \Omega_0({\bf k}_n,t)|b> \equiv
(\Omega^{ab}_{r}({\bf k}_n,t)+
i \Omega^{ab}_{i}({\bf k}_n,t)) e^{-i \omega_{ab}^R t}
\end{equation}
where
\begin{equation}
\omega_{ab}^R = \omega_a^R - \omega_b^R
\end{equation}
with $\omega_a^R$ being the energy of the a-th state of the uncoupled
reservoir.  Hermeticity of $\Omega$ implies that $\Omega_r^{ab}=\Omega_r^{ba}$
and $\Omega_i^{ab}=-\Omega_i^{ba}$.
We assume that the diagonal matrix elements are zero
\begin{equation}
\Omega_r^{aa}=0.
\label{omega0}
\end{equation}
This is equivalent to assuming that
there  is no time independent force exerted by the heat reservoir
on the system.
The only nontrivial term for which the statistical
expectation value needs to be evaluated is eq. (\ref{2term}).
The expectation value of the first term on the right hand side
of eq. (\ref{2term}) is zero due to eq. (\ref{omega0}).  However, we must
retain this term in its operator form since we will also want
products  of it in the final rate
equation when evaluating correlation functions.

For $L({\bf k}_n,t)$ in eq. (\ref{2term}), upon statistical averaging
we get
\begin{equation}
\frac{1}{N}
\sum_a \langle a | L({\bf k}_n,t) |a \rangle e^{-\omega^R_a/T}
= L^D_T({\bf k}_n,t)
+ L^C_T({\bf k}_n,t)
\end{equation}
where
\begin{equation}
L^D_T({\bf k}_n,t) = \frac{2\alpha^2}{\omega_n N L^3} \sum_{ab}
e^{-\omega^R_a/T} \sum_{n'}
\int_0^t dt' \sin \omega_n(t-t') \int_0^{t'} dt'' \sin \omega_{ab}^R (t'-t'')
\phi({\bf k}_{n'}, t'') \Omega^2_D(a,b,{\bf k}_n, {\bf k}_{n'}),
\label{a5}
\end{equation}
\begin{equation}
L^C_T({\bf k}_n,t) = -\frac{2\alpha^2}{\omega_n N L^3} \sum_{ab}
e^{-\omega^R_a/T} \sum_{n'}
\int_0^t dt' \sin \omega_n(t-t') \int_0^{t'} dt'' \cos \omega_{ab}^R (t'-t'')
\phi({\bf k}_{n'}, t'') \Omega^2_C(a,b,{\bf k}_n, {\bf k}_{n'})
\label{a5c}
\end{equation}
with
\begin{equation}
N \equiv \sum_a e^{-\omega_a^R/T},
\end{equation}
\begin{equation}
\Omega^2_D (a,b, {\bf k}_n, {\bf k}_{n'}) \equiv \Omega^{ab}_r({\bf k}_n)
\Omega^{ba}_r(-{\bf k}_{n'})
-\Omega^{ab}_i({\bf k}_{n}) \Omega_i^{ba}(-{\bf k}_{n'}),
\label{defD}
\end{equation}
and
\begin{equation}
\Omega^2_C (a,b, {\bf k}_n, {\bf k}_{n'}) \equiv \Omega^{ab}_r({\bf k}_n)
\Omega^{ba}_i(-{\bf k}_{n'})
+\Omega^{ab}_i({\bf k}_{n}) \Omega_r^{ba}(-{\bf k}_{n'}).
\label{defC}
\end{equation}

The energy levels in the heat reservoir are assumed to be closely spaced
so that we can replace the summation over the states by an integration as
\begin{equation}
\sum_{ab} \rightarrow \int_0^{\infty} \rho(\omega_a^R) d\omega_a^R
\int_0^{\infty} \rho(\omega_b^R) d\omega_b^R.
\end{equation}
We perform the above energy integration with the variables
\begin{eqnarray}
\omega_+^R &\equiv& \frac{\omega_a^R + \omega_b^R}{2}
\nonumber\\
\omega_-^R &\equiv& \omega_a^R - \omega_b^R,
\end{eqnarray}
so that
\begin{equation}
\int_0^{\infty} d\omega_a^R \int_0^{\infty} d \omega_b^R =
\int_0^{\infty} d\omega_-^R \int_{\frac{\omega_-^R}{2}}^{\infty} d \omega_+^R
+\int_{-\infty}^0 d\omega_-^R \int_{\frac{-\omega_-^R}{2}}^{\infty} d
\omega_+^R
\equiv \int d{\tilde \omega}_-^R \int d {\tilde \omega}_+^R.
\end{equation}
Eqs. (\ref{a5}) and (\ref{a5c}) then become respectively
\begin{eqnarray}
L^D_T({\bf k}_n, t) &=& \frac{-2\alpha^2}{\omega_n NL^3} \sum_{n'}
\int d{\tilde \omega}_-^R \int d {\tilde \omega}_+^R
\rho({\tilde \omega}_+^R+ \frac{{\tilde \omega}_-^R}{2})
\rho({\tilde \omega}_+^R - \frac{{\tilde \omega}_-^R}{2})
\nonumber\\
& &e^{-({\tilde \omega}_+^R+ \frac{{\tilde \omega}_-^R}{2})/T}
\Omega^2_D ({\tilde \omega}_+^R + \frac{{\tilde \omega}_-^R}{2},
{\tilde \omega}_+^R - \frac{{\tilde \omega}_-^R}{2},
{\bf k}_n, {\bf k}_{n'})
\nonumber\\
& & \int_0^t d\tau \left[ \frac{\cos({\tilde
\omega}_-^R+\omega_n)\frac{\tau}{2}
\sin({\tilde \omega}_-^R-\omega_n)\frac{\tau}{2}}
{{\tilde \omega}_-^R- \omega_n}
-\frac{\cos({\tilde \omega}_-^R - \omega_n)\frac{\tau}{2}
\sin({\tilde \omega}_-^R+\omega_n)\frac{\tau}{2}}
{{\tilde \omega}_-^R+ \omega_n} \right] \phi({\bf k}_{n'}, t-\tau)
\label{a61}
\end{eqnarray}
\begin{eqnarray}
L^C_T({\bf k}_n, t) &=& \frac{-2\alpha^2}{\omega_n NL^3} \sum_{n'}
\int d{\tilde \omega}_-^R \int d {\tilde \omega}_+^R
\rho({\tilde \omega}_+^R+ \frac{{\tilde \omega}_-^R}{2})
\rho({\tilde \omega}_+^R - \frac{{\tilde \omega}_-^R}{2})
\nonumber\\
& &e^{-({\tilde \omega}_+^R+ \frac{{\tilde \omega}_-^R}{2})/T}
\Omega^2_C ({\tilde \omega}_+^R + \frac{{\tilde \omega}_-^R}{2},
{\tilde \omega}_+^R - \frac{{\tilde \omega}_-^R}{2},
{\bf k}_n, {\bf k}_{n'})
\nonumber\\
& & \int_0^t d\tau \sin({\tilde \omega}_-^R+\omega_n)\frac{\tau}{2}
\sin({\tilde \omega}_-^R-\omega_n)\frac{\tau}{2}
\left[\frac{1}{{\tilde \omega}_-^R- \omega_n}
-\frac{1}
{{\tilde \omega}_-^R+ \omega_n} \right] \phi({\bf k}_{n'}, t-\tau).
\label{a62}
\end{eqnarray}
Define
\begin{equation}
B^I(\omega, {\bf k}_n, {\bf k}_{n'}) \equiv \int_0^{\infty} d \omega '
\rho(\omega'+ \omega) \rho(\omega ')
\Omega^2_I(\omega'+\omega, \omega ', {\bf k}_n, {\bf k}_{n'})
e^{-\omega'/T},
\label{defB}
\end{equation}
where $I=C$ or $I=D$.  We assume $B^I$ is a smooth function
with respect to $\omega$.  Identifying this term in eqs. (\ref{a61})
and (\ref{a62}) we observe that due to the oscillatory factor
the main contribution to the
${\tilde \omega}_-^R$ integral is from
${\tilde \omega}_-^R = \pm \omega_n$.  Evaluating the
${\tilde \omega}_-^R$ integration under this approximation,
we obtain
\begin{equation}
L^D_T({\bf k}_n, t) = \sum_{n'} \Gamma^D_{nn'}\int_0^t dt'
\phi({\bf k}_{n'}, t') \cos \omega_n (t-t')
\end{equation}
\begin{equation}
L^C_T({\bf k}_n, t) = \sum_{n'} \Gamma^C_{nn'}\int_0^t dt'
\phi({\bf k}_{n'}, t') \sin \omega_n (t-t')
\end{equation}
where
\begin{equation}
\Gamma^I_{nn'} \equiv \frac{\pi \alpha^2}{\omega_n N L^3}
(1-e^{-\omega_n/T}) B^I(\omega_n,{\bf k}_n, {\bf k}_{n'})
\label{defgam}
\end{equation}
with $I=C,D$.
Substituting this into eq. (\ref{A1}) we obtain the equation of motion
\begin{equation}
\ddot{\phi}({\bf k}_n,t) +\omega_n^2 \phi({\bf k}_n,t)
+\sum_{n'} \Gamma^D_{nn'} {\dot \phi}({\bf k}_{n'},t)
+\omega_n \sum_{n'} \Gamma^C_{nn'} \phi({\bf k}_{n'},t)
+\frac{\delta V}{\delta \phi (-{\bf k}_n,t)} = \eta({\bf k}_n,t),
\end{equation}
where $\eta({\bf k}_n,t) \equiv - \alpha \Omega_0({\bf k}_n,t)$.
Hereafter we will assume that the uncoupled reservoir system is translationally
invariant so that
\begin{equation}
\langle \langle \Omega_0({\bf k}_n,t)
\Omega_0({\bf k}_{n'},t') \rangle \rangle_T = \delta_{n, - n'}
\langle \langle \Omega_0({\bf k}_n,t)
\Omega_0(-{\bf k}_n,t) \rangle \rangle_T.
\end{equation}
The statistical properties of $\eta({\bf k}_n,t)$ are
\begin{equation}
\langle \langle \eta({\bf k}_n,t) \rangle \rangle_T=0
\end{equation}
\begin{equation}
\langle \langle \eta({\bf k}_n,t) \eta({\bf k}_{n'},t') \rangle \rangle_T=
\delta_{n, -n'} \frac{\omega_n \Gamma^D_{n,n} L^3}{\pi}
\left( \frac{iP}{t-t'} + \pi \delta (t-t')
\coth \left(\frac{\omega_n}{2T} \right) \right).
\label{etaeta}
\end{equation}
The first property follows from eq. (\ref{omega0}).  The second follows from
eqs. (\ref{defD}), (\ref{defC}), (\ref{defB}), and (\ref{defgam})
and using
\begin{equation}
\int_0^{\infty} d\omega e^{i\omega t} = \frac{iP}{t} + \pi \delta(t).
\end{equation}
In the high temperature limit, eq. (\ref{etaeta}) becomes
\begin{equation}
\langle \langle \eta({\bf k}_n,t) \eta(-{\bf k}_{n},t')
\rangle \rangle_{T \rightarrow \infty} \rightarrow
2L^3 \Gamma^D_{n, n} T \delta(t-t'),
\end{equation}
which agrees with eq. (\ref{fdvalue}).

\end{document}